\documentclass[preprint]{vgtc}               

\ifpdf
  \pdfoutput=1\relax                   
  \usepackage{graphicx}                
  \DeclareGraphicsExtensions{.pdf,.png,.jpg,.jpeg} 
\else
  \usepackage{graphicx}                
  \DeclareGraphicsExtensions{.eps}     
\fi%

\graphicspath{{figures/}{pictures/}{images/}{./}} 

\usepackage{microtype}                 
\PassOptionsToPackage{warn}{textcomp}  
\usepackage{textcomp}                  
\usepackage{mathptmx}                  
\usepackage{times}                     
\usepackage{cite}                      
\usepackage{tabu}                      
\usepackage{booktabs}                  

\onlineid{0}

\vgtccategory{Application or design study}

\vgtcinsertpkg

\ieeedoi{10.1109/VIS54862.2022.00034}

\usepackage{caption}
\usepackage{subcaption}
\usepackage{enumitem}

\title{Color Coding of Large Value Ranges Applied to Meteorological Data}

\author{Daniel Braun\thanks{e-mail: braun@cs.uni-koeln.de}\\ %
       \parbox{1.4in}{\scriptsize \centering University of Cologne \\ Department of Mathematics and Computer Science} %
\and Kerstin Ebell\thanks{e-mail: kebell@meteo.uni-koeln.de}\\ %
     \parbox{1.4in}{\scriptsize \centering University of Cologne \\ Institute for Geophysics and Meteorology} %
\and Vera Schemann\thanks{e-mail: schemann@meteo.uni-koeln.de}\\ %
     \parbox{1.4in}{\scriptsize \centering University of Cologne \\ Institute for Geophysics and Meteorology} %
\and Laura Pelchmann\thanks{e-mail: pelchmann@cs.uni-koeln.de}\\ %
     \parbox{1.4in}{\scriptsize \centering University of Cologne \\ Department of Mathematics and Computer Science} %
\and Susanne Crewell \thanks{e-mail: susanne.crewell@uni-koeln.de}\\ %
     \parbox{1.4in}{\scriptsize \centering University of Cologne \\ Institute for Geophysics and Meteorology} %
\and Rita Borgo\thanks{e-mail: rita.borgo@kcl.ac.uk}\\ %
     \parbox{1.4in}{\scriptsize \centering King's College London \\ Department of Computer Science} %
\and Tatiana von Landesberger\thanks{e-mail: landesberger@cs.uni-koeln.de}\\ %
     \parbox{1.4in}{\scriptsize \centering University of Cologne \\ Department of Mathematics and Computer Science}} %

\teaser{
  \centering
     \begin{subfigure}[tb]{0.32\textwidth}
         \centering
         \includegraphics[width=\textwidth]{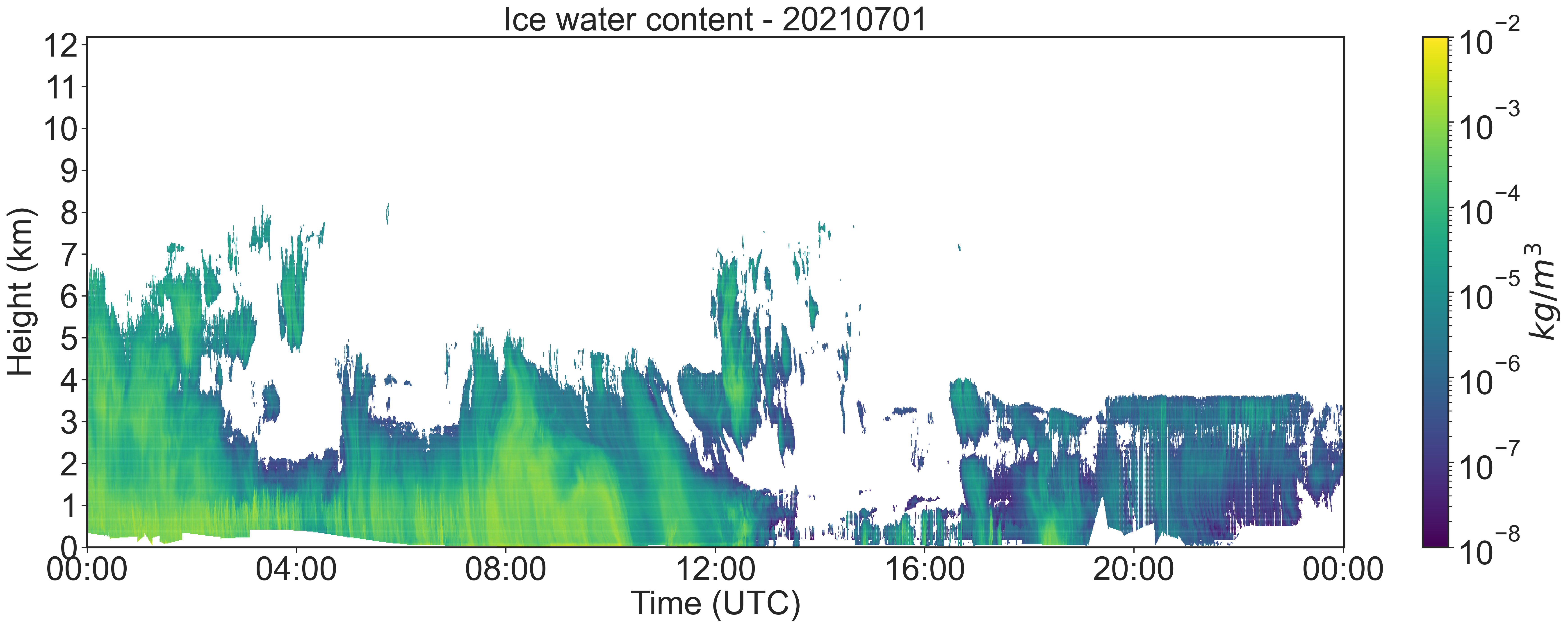}
         \caption{Viridis}
         \label{fig:scat_v}
     \end{subfigure}
     \hfill
     \begin{subfigure}[tb]{0.32\textwidth}
         \centering
         \includegraphics[width=\textwidth]{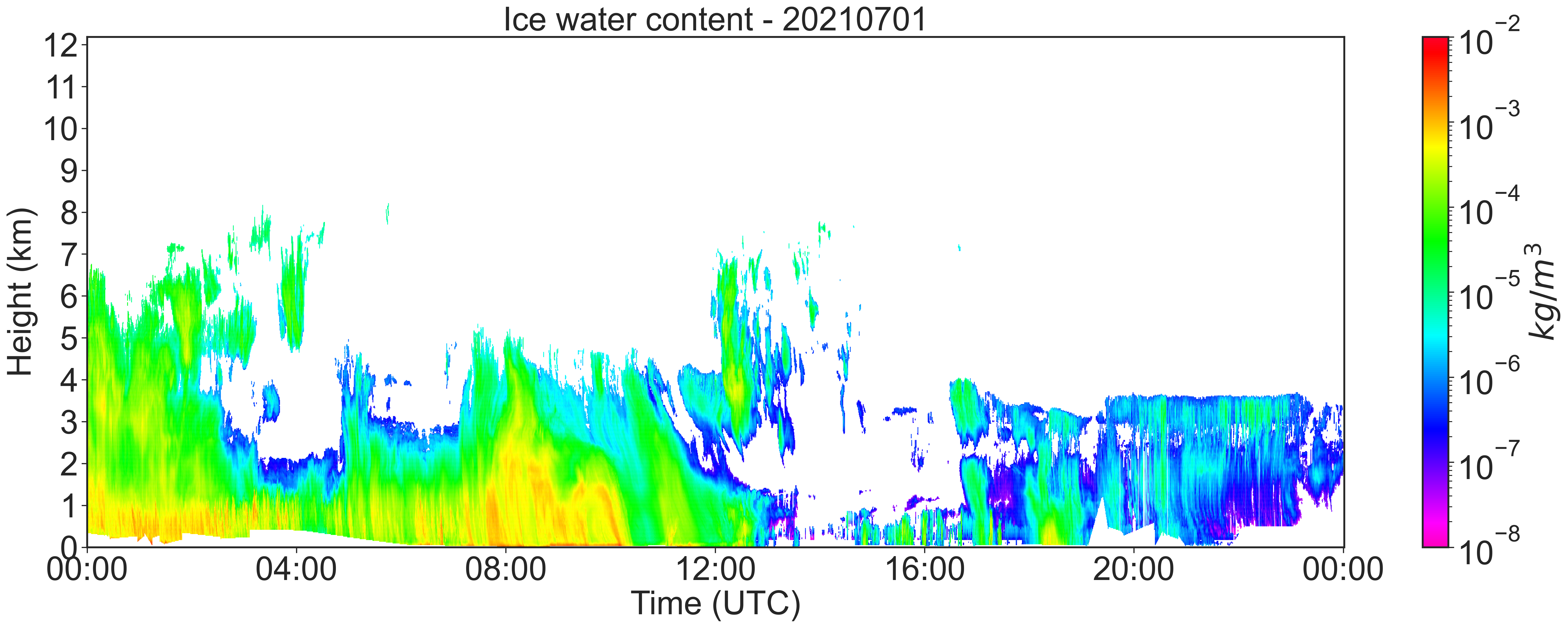}
         \caption{Rainbow}
         \label{fig:scat_r}
     \end{subfigure}
     \hfill
     \begin{subfigure}[tb]{0.32\textwidth}
         \centering
         \includegraphics[width=\textwidth]{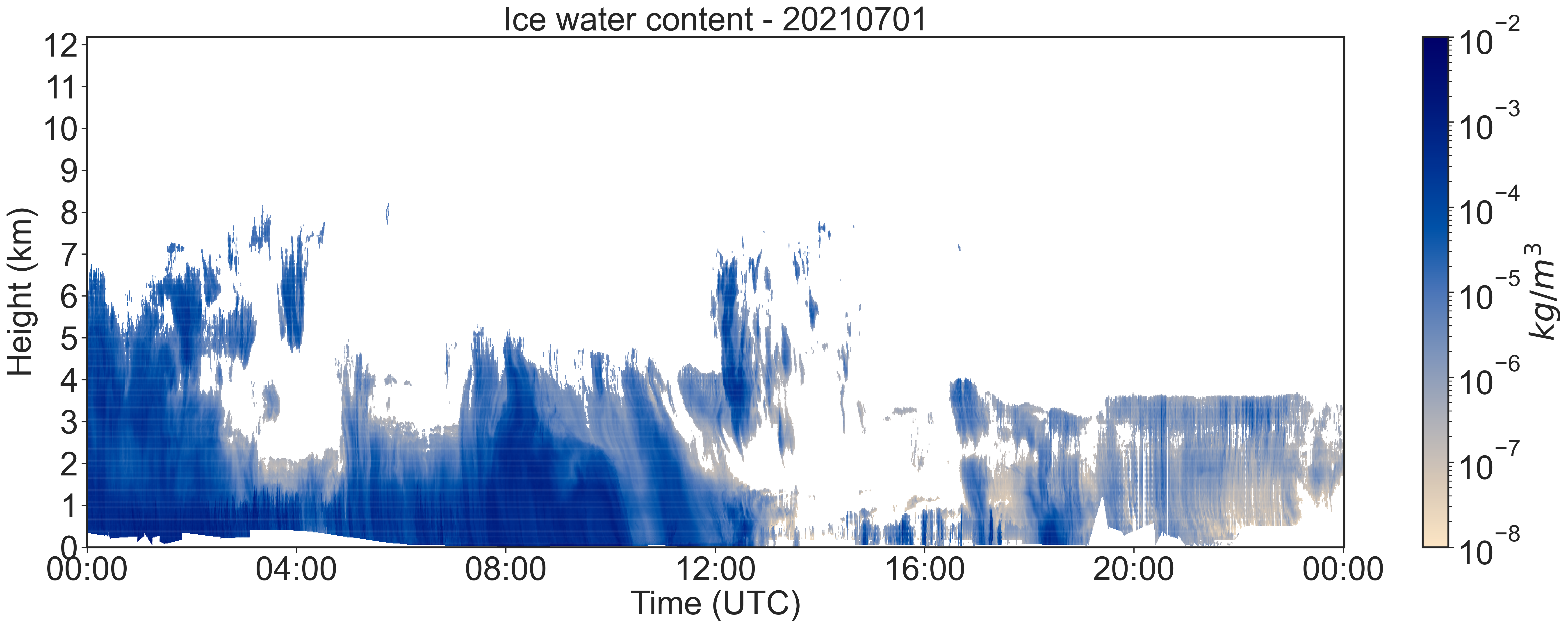}
         \caption{ColorCrafter \cite{Smart.2020}}
         \label{fig:scat_c}
     \end{subfigure}
     
     \begin{subfigure}[tb]{0.49\textwidth}
         \centering
         \includegraphics[width=\textwidth]{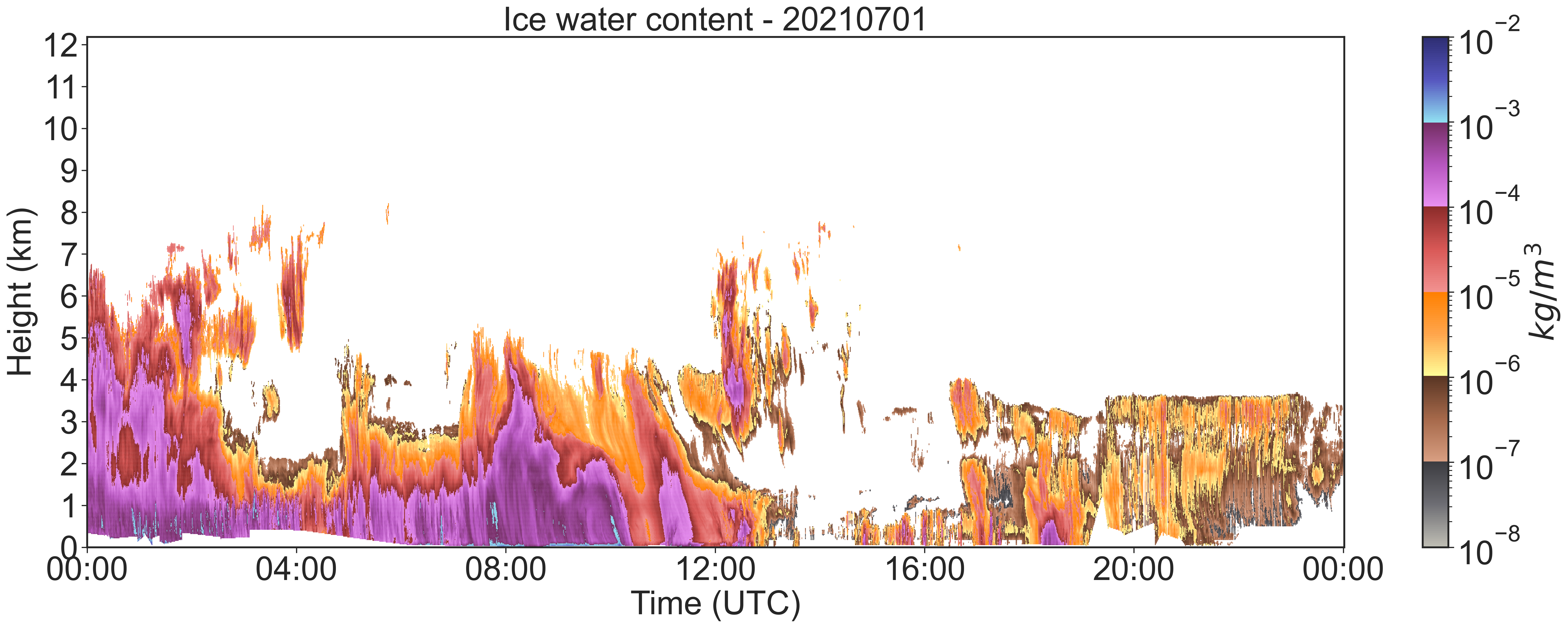}
         \caption{Order of magnitude colors (OMC)}
         \label{fig:scat_OMC}
     \end{subfigure}%
     \hfill
     \begin{subfigure}[tb]{0.49\textwidth}
         \centering
         \includegraphics[width=\textwidth]{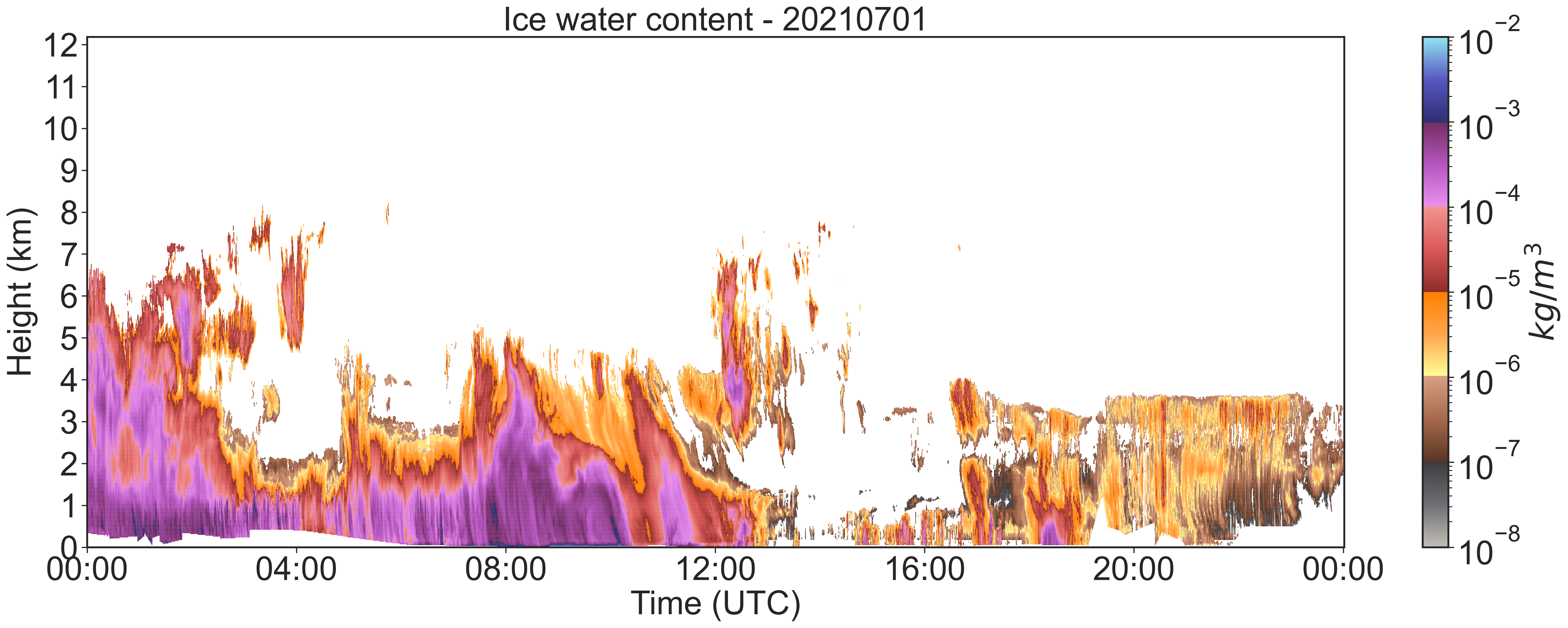}
         \caption{OMC smoothed lightness ($\mbox{OMC}_{sl}$)}
         \label{fig:scat_OMCsl}
     \end{subfigure}
        \caption{Scatterplots of meteorological data for the ice water content in clouds (time height series) featuring large value ranges encoded using the color schemes Viridis, Rainbow, ColorCrafter and two versions of our \textit{order of magnitude colors} design.}
        \label{fig:teaser}
}

\abstract{This paper presents a novel color scheme designed to address the challenge of visualizing data series with large value ranges, where scale transformation provides limited support. We focus on meteorological data, where the presence of large value ranges is common. We apply our approach to meteorological scatterplots, as one of the most common plots used in this domain area. Our approach leverages the numerical representation of mantissa and exponent of the values to guide the design of novel ``nested'' color schemes, able to emphasize differences between magnitudes. Our user study evaluates the new designs, the state of the art color scales and representative color schemes used in the analysis of meteorological data: ColorCrafter, Viridis, and Rainbow. We assess accuracy, time and confidence in the context of discrimination (comparison) and interpretation (reading) tasks. Our proposed color scheme significantly outperforms the others in interpretation tasks, while showing comparable performances in discrimination tasks.}

\CCScatlist{
 \CCScatTwelve{Color Coding}{Perception}{Large Value Ranges}{User study}
}

\begin{document}

\maketitle

\section{Introduction}

Data with large value ranges are data sets whose values contain several different exponents using the scientific notation ($v=m \cdot 10^{e}$)~\cite{Hohn.2020}. One example is meteorological data on ice water content in clouds, whose exponents vary between minus eight and minus two.
The data includes three variables: time, height and ice water content. Mapping the values to position is not possible in a 2D-visualization. 
The current meteorological standard is to visualize the data using a scatterplot with the values encoded by a logarithmic scaled colormap~\cite{meteo, cloudnet}.
The meteorologists call this \textit{time-height series}. The challenge is to encode such a large value range on a color scale and a limited number of pixels without loss of information.

The common colormaps used for meteorological scatterplots are \textit{Viridis} (\autoref{fig:scat_v})~\cite{cloudnet} and \textit{Rainbow} (\autoref{fig:scat_r})~\cite{meteo}. We present a novel color scheme to visualize data with large value ranges: the \textit{order of magnitude colors} (OMC), which uses a seperate representation of the mantissa and exponents of the data (\autoref{fig:scat_OMC}) for an easier classification of the orders of magnitude. It uses one hue for each exponent and linear gradient of brightness within for magnitude. Additionally a variation of this design is shown (\autoref{fig:scat_OMCsl}).

We compare our approach with the currently used color scales as well as a state-of-the-art color scheme generated via the tool \textit{ColorCrafter}~\cite{Smart.2020} (starting color: blue) (\autoref{fig:scat_c}) in an empirical user study. The results show that our new color scheme has comparable results for comparison tasks and works significantly better for reading values than the other colormaps. We already received requests from meteorologists for the use of our design.

\section{Related Work}

\paragraph{Color Perception}
Golebiowska and Coltekin~\cite{Golebiowska.2020} explored the perceptual differences from rainbow to sequential color scales. Their study showed that the sequential colormaps performed better in comparing values and recognizing general patterns, whereas the rainbow scale supports the reading of specific details. We also compare these two types of color schemes for the investigated tasks.

The pros and cons of the rainbow color scheme have been widely discussed in literature. Rogowitz and Treinish~\cite{Rogowitz.1998} criticize the Rainbow-colormap for the fact that the boundaries of the different colors can be perceived as boundaries in the data. This is countered by the findings of Reda and Szafir \cite{Reda.2021} and Reda et al. \cite{Reda.2021b} that the more unique colors represented, the better the plot.

There are a variety of tools that can be used to generate and optimize color scales.
The \textit{ColorCrafter}-tool~\cite{Smart.2020} creates algorithmically generated sequential color schemes for given parameters. These colormaps work well for the most quantitative data, but hide details when visualizing data with large value ranges. Another software is the~\textit{CCC}-tool, which allows to create completely new color schemes and to optimize given color scales~\cite{Nardini.2021}.

A data-dependent adaptation of color schemes based on statistical properties~\cite{SchulzeWollgast.2005, Tominski.2008} would lead to incomparability of daily views. Thus, we disregard them.

\paragraph{Visualization of Large Value Ranges}
Most of the research work on large value ranges introduced novel visualization types for two-dimensional data. The methods for bar charts presented by Hlawatsch et al.~\cite{Hlawatsch.2013}, Borgo et al.~\cite{Borgo.2014} and Höhn et al.~\cite{Hohn.2020} are not applicable for our use case. They provide inspiration for our work - the separate representation of the exponent and the mantissa.

\paragraph{Taxonomy of Tasks}

Most task types can be divided into high- and low-level tasks~\cite{Brehmer.2013}. There are several papers that deal with low-level tasks and introduce their own taxonomies of tasks \cite{Amar.2005, Quadri.2021, Sarikaya.2018, Valiati.2006}. In our user study, we use two low-level tasks that appear in all of the taxonomies to evaluate the color scales.

\section{Color Scheme Design}

\begin{figure}
\centering
\hspace*{.059\linewidth}\includegraphics[width=0.92\columnwidth]{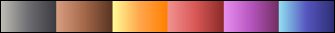}
\hspace*{.018\linewidth}\includegraphics[width=0.98\columnwidth]{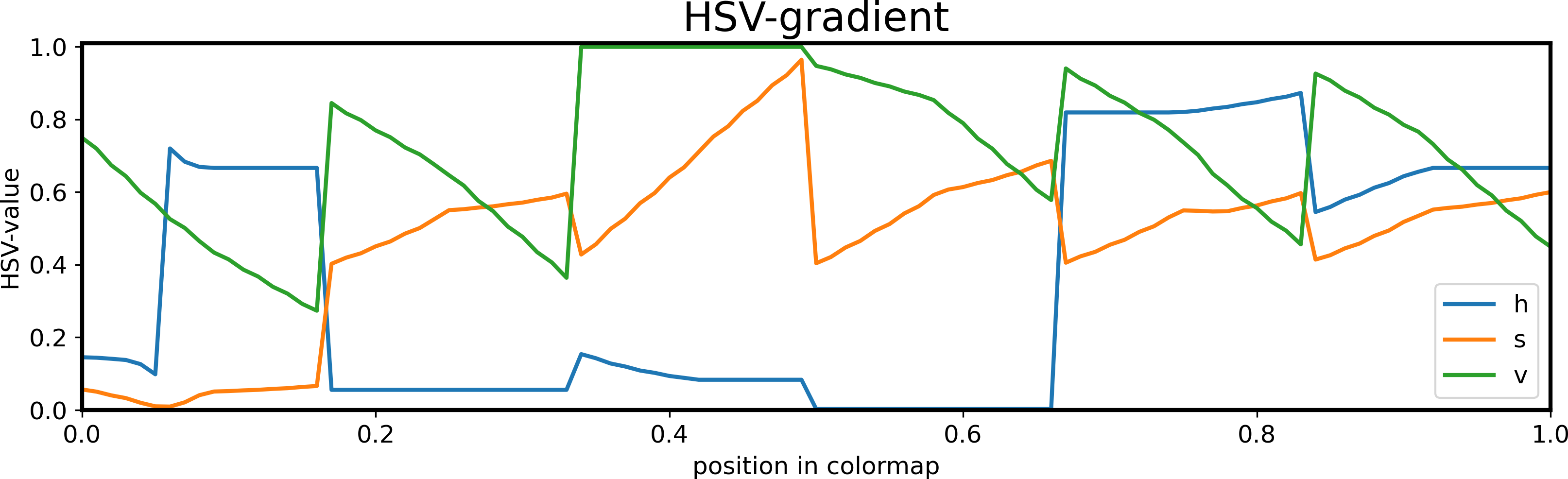}
\includegraphics[width=\columnwidth]{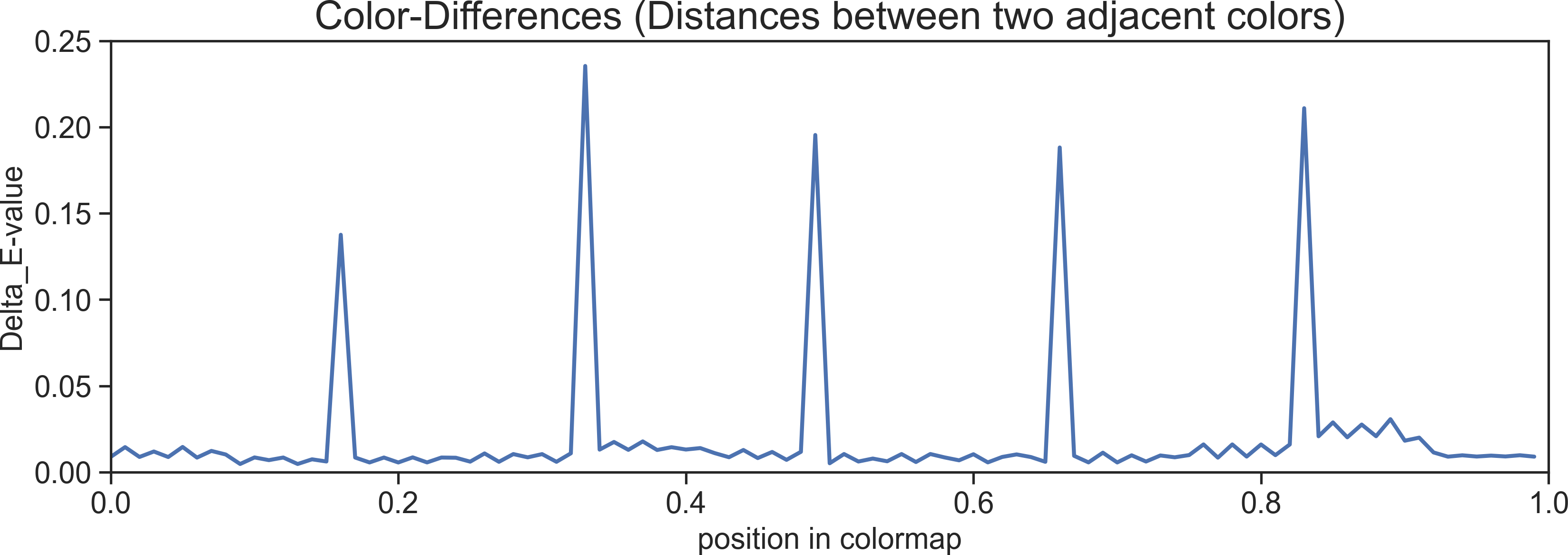}
\caption{\textit{Order of magnitude colors} color scheme.}
\label{fig:OMC}
\end{figure}

We present a novel color scheme named \textit{order of magnitude colors} (\autoref{fig:OMC}). This design is inspired by the scientific notation of numbers and is created in the \textit{CCC}-tool~\cite{Nardini.2021}. Inspired by the idea of the recent approaches~\cite{Hlawatsch.2013, Borgo.2014, Hohn.2020}, we use the two parts mantissa $m$ and exponent $e$ of a value $v$ (so that $v=m \cdot 10^{e}$) for color coding. Every exponent given in the data is mapped to another hue. Within an exponent, the mantissa is mapped to a perceptually linear sequential scale of the respective hue.  

During the research process, we created and tried many different color scales to the meteorological data (see the supplementary material). The color selection as well as the data and the study tasks were discussed and agreed upon in several focus groups organized together with the meteorologists. In the final design, the colors were chosen through a manual process so that the color distances between adjacent colors are approximately even. Color blindness would be an additional criteria for further research. We made the HSV-gradient for the different hues as smooth and equal as possible (see \autoref{fig:OMC}). To achieve a better color nameability~\cite{Reda.2021, Reda.2021b}, we chose unique hues instead of different saturation tones of one specific hue. 

\begin{figure}
\centering
\hspace*{.059\linewidth}\includegraphics[width=0.92\columnwidth]{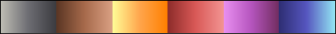}
\hspace*{.02\linewidth}\includegraphics[width=0.98\columnwidth]{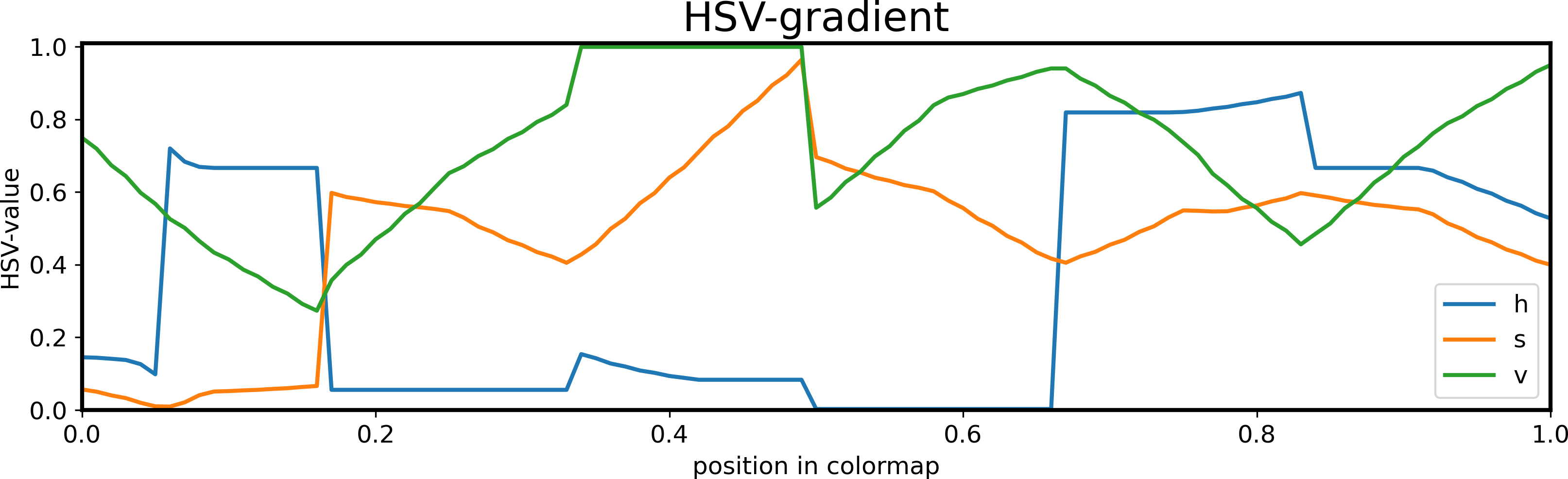}
\includegraphics[width=\columnwidth]{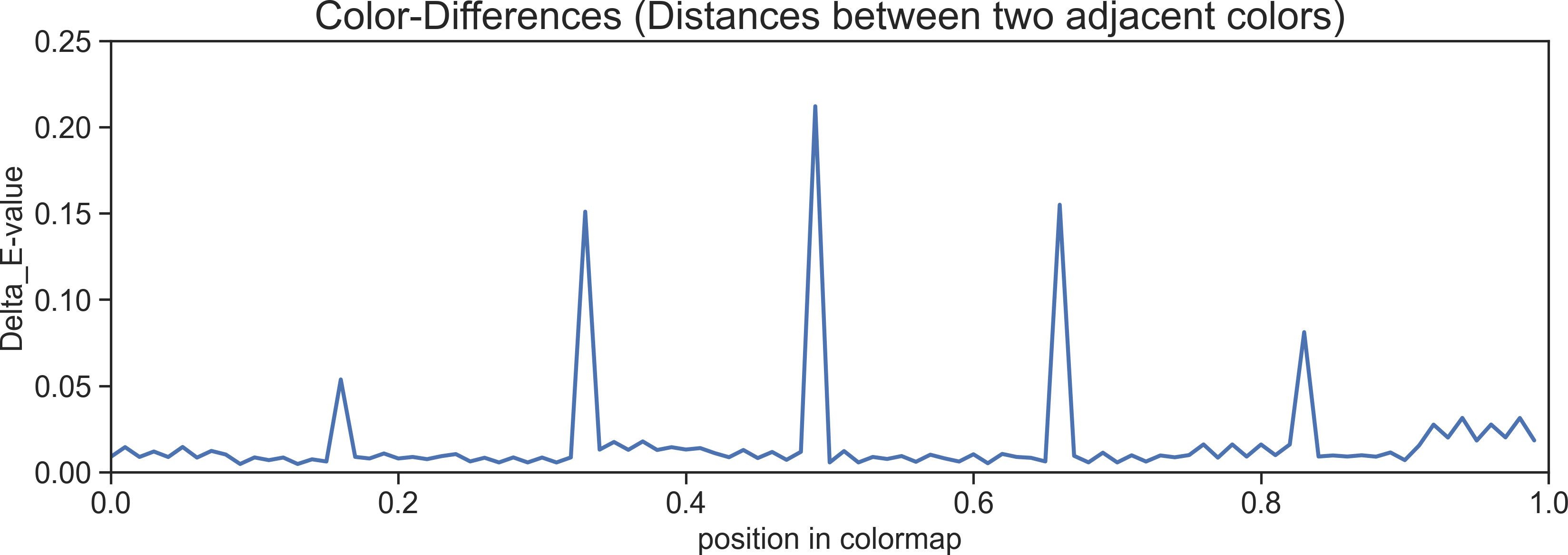}
\caption{\textit{OMC smoothed lightness} color scheme.}
\label{fig:OMCsl}
\end{figure}

Boundaries in colors are easily perceived as changes in the exponents in the data~\cite{Rogowitz.1998}. This is deemed appropriate, as our colormap describes the changes in magnitudes. It helps the meteorologists looking at changes in the data. Nevertheless, we also created a variation of the OMC-colormap that reduces the color distances (i.e. smaller DeltaE-values~\cite{McDonald.1995}) (see \autoref{fig:OMC} and \autoref{fig:OMCsl}). In this version, we flipped every second sequential color scale so that we have a changing lightness direction comparing two neighboring color scales: The \textit{OMC smoothed lightness} (\autoref{fig:OMCsl}). 

\section{Evaluation}

\textit{Viridis} and \textit{Rainbow} are common color schemes for visualizing meteorological data. Therefore, we evaluate these two, a state-of-the-art colormap created by \textit{ColorCrafter}~\cite{Smart.2020} and our two new designs - the OMC and $\mbox{OMC}_{sl}$ - in an exploratory user study.

\paragraph{Data}

The data used for the study were real meteorological data sets provided by the Institute for Geophysics and Meteorology at the University of Cologne. Every set of data contains one measurement day with the variables \textit{time} (0 to 24 hours), \textit{height} (0 to 12 kilometers) and the \textit{ice water content} in clouds. The ice water content spans a range of $10^{-8}$ to $10^{-2}$. Due to the large value ranges, the colormaps were scaled logarithmic.

To avoid a learning effect, each question of the study contained visualizations of different measurement data. For consistency, we used data sets with similar properties, e.g., an equal value range.

\paragraph{Tasks}

In line with literature and after consultation with the meteorologists, we decided to focus on four main tasks, divided into reading~\cite{Amar.2005, Sarikaya.2018, Quadri.2021, Brehmer.2013, Valiati.2006} and comparison~\cite{Quadri.2021, Sarikaya.2018, Brehmer.2013, Valiati.2006} tasks:

\begin{itemize}[noitemsep]
\item \textit{Reading Tasks}: 
\begin{itemize}[noitemsep, nolistsep]
    \item \textit{Extrema}: name the maximum and minimum exponent of a given day
    \item \textit{Value:} specify a value range for the marked region of data points
\end{itemize}
\item \textit{Comparison Tasks}: 
\begin{itemize}[noitemsep, nolistsep]
    \item \textit{Extrema}: compare two measurement days and decide which one contains the global extremum
    \item \textit{Value:} compare the values of two marked regions of data points in one day
\end{itemize}
\end{itemize}

We manually selected the regions to ensure that the tasks for each colormap have the same level of difficulty. The comparison tasks are single choice, multiple options questions. In the reading tasks, the participants had to insert their answered numbers manually. Additionally, we asked the participants about their confidence in the given answers of the reading tasks.

\paragraph{Experimental Setting and Procedure}

A total of 53 participants (36 male, 13 female and 4 prefer not to say) took part in the study. The age was distributed from 20 up to 60, but the majority of the participants were between 20 and 40 years (88\%). We filtered out 5 participants, who did not correctly answer to the Ishihara tests for color blindness~\cite{Clark.1924} (this was necessary because the rainbow scale is not suitable for color blindness). 

The expertise of the participants was broad and ranged from a degree in mathematics (25\%) to a degree in physics (13\%) to a degree in computer science (10\%) and others (4\%), with the majority of the participants having a degree in meteorology (48\%).

The study was conducted online and was set up with \textit{LimeSurvey}~\cite{LimeSurvey.2022}. There was one task per page and the participants had to click on a button manually to get to the next page, so we could store the given answers and the response time per task. The visualizations used had a resolution of $1291 x 500$px. All of the participants used a computer screen with a size of at least 13''.

The processing time of the study was approximately 30 minutes. All participants had to solve the same tasks. There was one task for every of the five designs for every type of task, i.e. in total we had $5[color scales] \times 4[tasks]=20$ trials.

After a short training phase introducing the visualization type and the types of tasks, the sorting of the tasks was randomized to reduce a learning effect. At the end of the study the participants were asked to give feedback. Study documentation is in annex.

\subsection{Analysis} 

The analysis of the results is performed in \textit{R}~\cite{R.2021}. For every task of the study, we measure accuracy and response time. In the reading tasks, the specified confidence of the participants is also examined (on a Likert scale from 1 -- very unsure to 5 -- very confident). 

\paragraph{Correctness measurement}

For every task, we compare the number of correct answers to the number of incorrect answers. In the extrema reading task, both exponents have to be correct to be considered as a correct answer. For value reading an answer is correct, if the values of the marked region lie in the answered range.

In addition, the sizes of the specified ranges in the reading value task are compared. To exclude logarithmic scaling, the size of the ranges is calculated using~\autoref{eq:size}:
\begin{equation}
    \mbox{range size} = \frac{(Exp_{max}-Exp_{min}) \cdot 10 + (Mant_{max}-Mant_{min})}{10}
    \label{eq:size}
\end{equation}

\paragraph{Significance Tests}

To perform our analysis, a three-stage significant test for each task was used. Since we could not assume that the data is normally distributed, we first ran a Shapiro-Wilk test on the given answers and response times, with the result that none of the data is normally distributed.

In the second stage of the analysis, we used a $\chi^{2}$-test (and a Fisher-test if the frequencies are too low) to investigate significance in the amounts of correct and wrong answers. For the quantitative data like the range sizes or response time, we used the Kruskal-Wallis test.

In the third stage, as post-hoc analysis we used a Wilcoxon-test respectively a $\chi^{2}$-test for pairwise comparison of the color schemes for tasks for which significance was found. \\All tests were performed with the significance level $\alpha=0.05$.

\subsection{Results}

\begin{figure*}[!htbp]
 \centering
 \includegraphics[width=\textwidth]{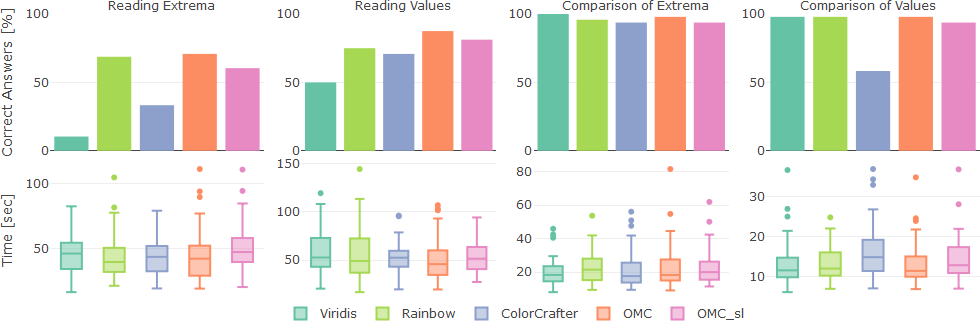}
 \caption{Percentages of correct answers (the more the better) and response times in seconds (the less the better) for tasks and colormaps.}
 \label{fig:result}
\end{figure*}

\autoref{fig:result} shows a summary of the results including the percentages of correct answers for the different colormaps in the four tasks and the response time of the participants.

\paragraph{Reading Extrema}

Our OMC design (71\% correct answers) and the rainbow color scheme (69\%) perform best. The $\chi^{2}$-test shows a significant dependence between the color scales and the amount of correct/ false answers ($\chi^{2} = 52.466$, $\mbox{p-value} = 1.102\mathrm{e}{-10}$). The post-hoc pairwise test confirms that OMC and Rainbow deliver significantly better results than ColorCrafter (33\%) and Viridis (10\%) (\autoref{tab:overview}). $\mbox{OMC}_{sl}$ is third best with 60\% of correct answers.

\begin{table}[tb]
\begin{tabular}[tb]{l|c|c|c|c}
\textbf{p-value} & \textit{OMC} & \textit{$OMC_{sl}$} & \textit{ColorCrafter} & \textit{Rainbow} \\
\hline
\textit{$OMC_{sl}$} & 0.390 & - & - & - \\
\hline
\textit{ColorCrafter} & \underline{0.001} & \underline{0.014} & - & - \\
\hline
\textit{Rainbow} & 1.000 & 0.522 & \underline{0.001} & - \\
\hline
\textit{Viridis} & \underline{0.000} & \underline{0.000} & \underline{0.014} & \underline{0.000} \\
\end{tabular}
\caption{p-values of the pairwise $\chi^{2}$-test for extrema reading.}
\label{tab:overview}
\end{table}

\paragraph{Reading Values}

The OMC designs perform best with respect to the number of correct answers. The $\chi^{2}$-test indicates significance ($\chi^{2} = 19.833$, $\mbox{p-value} = 0.001$). Due to the pairwise test, only the Viridis colormap is significantly worse than other color schemes.

\begin{figure}[tb]
 \centering
 \includegraphics[width=\columnwidth]{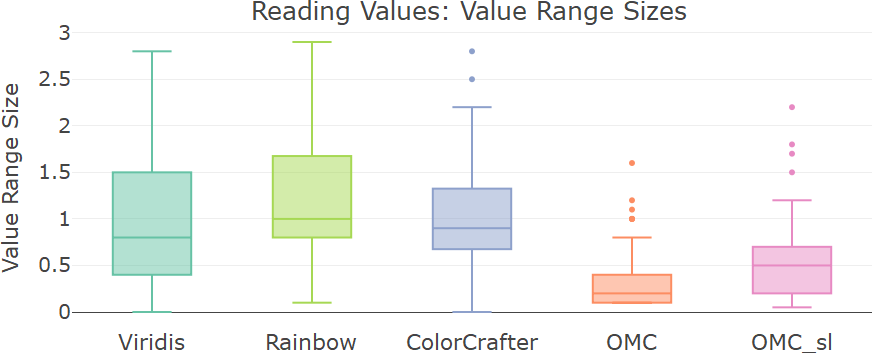}
 \caption{Sizes of the value ranges answered for value reading.}
 \label{fig:ident_range}
\end{figure}

The positive effect of the OMC colormap can be seen better in the comparison of the given range sizes (\autoref{fig:ident_range}). The OMC colormap has by far the smallest range size ($\mbox{mean}=0.36$), followed by $\mbox{OMC}_{sl}$ ($0.68$). Viridis ($0.91$), ColorCrafter ($0.99$) and Rainbow ($1.15$) are at a similar level. The Kruskal-Wallis-test confirms the significant differences in the means ($\chi^{2} = 50.518$, $\mbox{p-value} = 2.814\mathrm{e}{-10}$). The pairwise Wilcoxon-test shows, that the OMC color scale leads to significantly smaller range sizes (i.e. a better identification of the searched values) than all others (\autoref{tab:ident}). $\mbox{OMC}_{sl}$ also performs significantly better than ColorCrafter and Rainbow. This can be attributed to our design approach, where different hues enable fast recognizing of magnitudes.

\begin{table}[tb]
\begin{tabular}[tb]{l|c|c|c|c}
\textbf{p-value} & \textit{OMC} & \textit{$OMC_{sl}$} & \textit{ColorCrafter} & \textit{Rainbow} \\
\hline
\textit{$OMC_{sl}$} & \underline{0.000} & - & - & - \\
\hline
\textit{ColorCrafter} & \underline{0.000} & \underline{0.003} & - & - \\
\hline
\textit{Rainbow} & \underline{0.000} & \underline{0.000} & 0.166 & - \\
\hline
\textit{Viridis} & \underline{0.000} & 0.051 & 0.511 & 0.063 \\
\end{tabular}
\caption{p-values of the pairwise Wilcoxon-test for value reading.}
\label{tab:ident}
\end{table}

\paragraph{Comparison of Extrema}

Very few errors were made. ColorCrafter (3 errors) and $\mbox{OMC}_{sl}$ (3) have the most errors, followed by Rainbow (2) and OMC (1). Viridis even produced no error at all. Due to the low number of errors, no significant dependence can be determined using Fisher's exact test ($\mbox{p-value} = 0.471$).

\paragraph{Comparison of Values}

ColorCrafter performs worst for this task (58\% correct answers). $\mbox{OMC}_{sl}$ has 94\% of correct answers, while the other three even have 98\%. Due to the $\chi^{2}$-test there is a significant effect ($\chi^{2} = 66.679$, $\mbox{p-value} = 1.332\mathrm{e}{-9}$). The post-hoc pairwise Fisher-test shows that ColorCrafter leads to significantly more errors than all other color scales (\autoref{tab:comp_val}).

\subsection{Extended Analysis and Results}


\paragraph{Confidence}
For the reading extrema task, the analysis of the participants' confidence shows that the sequential color schemes (Viridis, ColorCrafter) lead to significantly more uncertainty in the answers than the multicolored schemes (OMC, $\mbox{OMC}_{sl}$, Rainbow).

For reading value task, the participants were significantly more confident with our OMC color scale than with all others, both for the exponents and the mantissas of the specified range.

\begin{table}[tb]
\begin{tabular}[tb]{l|c|c|c|c}
\textbf{p-value} & \textit{OMC} & \textit{$OMC_{sl}$} & \textit{ColorCrafter} & \textit{Rainbow} \\
\hline
\textit{$OMC_{sl}$} & 0.617 & - & - & - \\
\hline
\textit{ColorCrafter} & \underline{0.000} & \underline{0.000} & - & - \\
\hline
\textit{Rainbow} & 1.000 & 0.617 & \underline{0.000} & - \\
\hline
\textit{Viridis} & 1.000 & 0.617 & \underline{0.000} & 1.000 \\
\end{tabular}
\caption{p-values of the pairwise Fisher-test for value comparison}
\label{tab:comp_val}
\end{table}

\paragraph{Time}
Only for the value comparison task, the Kruskal-Wallis-test shows a significant mean effect in the response time ($\chi^{2} = 13.398$, $\mbox{p-value} = 0.009$). As the pairwise Wilcoxon-test shows, participants used significantly more time with ColorCrafter than with all others. Notably, the participants did not need any additional time to understand our new \textit{order of magnitude colors} design.

\paragraph{Free Text}
31 participants gave free text feedback. Almost all of them were positive regarding our OMC color scale. We already received requests to use our design. It was stated that the color scheme supports value comparison and identification by its clear borders: "The color scale with different colors between every power of 10 is the one that allowed me to better identify the values and the differences between points." Suggestions for improvement were the color selection and the perception of qualitative patterns.

The sequential colormaps were perceived as more aesthetically pleasing. Rainbow and ColorCrafter got the most negative feedback, especially for their color gradients.

\paragraph{Expertise}
Comparing the results of participants with a meteorological background to the others, there was no significant difference in the correctness of their answers and in the response time. This indicates that our color scheme can be used broadly, without specific expertise or background.

\paragraph{Limitations}
There are some limitations to our color scheme. The approach is not suitable for comparing values that are close to the borders of the exponents or for data containing positive and negative values. Additionally, further experiment may be needed to assess the scalability of the proposed scheme for value ranges with different exponent distributions. In order not to make the user study too extensive, we focused on a subset of tasks most relevant to our collaborators. Therefore, some types of task could not be dealt with sufficiency. For example, an evaluation of the OMC colormap for high level tasks would be interesting.

\section{Conclusion and Future Work}

We presented a new color coding approach to visualize data featuring large value ranges with the application to meteorological data. The empirical study has shown that our order of magnitude colors design performs very well for all tasks examined. It has no significant difference to the best performing colormaps in the comparison tasks and improves the accuracy of value identification significantly. The strict borders of our color scheme support the perception of the different orders of magnitude.

In summary, our results suggest the following ranking for each proposed design:
\begin{itemize}[noitemsep, nolistsep]
    \item \textit{Reading Extrema}: \\ \textbf{OMC} $\succeq$ Rainbow $\succeq$ $\mbox{OMC}_{sl}$ $\succ$ ColorCrafter $\succ$ Viridis
    \item \textit{Reading Values}: \\ \textbf{OMC} $\succ$ $\mbox{OMC}_{sl}$ $\succeq$ Viridis $\succeq$ ColorCrafter $\succeq$ Rainbow
    \item \textit{Comparison of Extrema}: \\ Viridis $\succeq$ \textbf{OMC} $\succeq$ Rainbow $\succeq$ $\mbox{OMC}_{sl}$ $\sim$ ColorCrafter
    \item \textit{Comparison of Values}: \\ \textbf{OMC} $\sim$ Viridis $\sim$ Rainbow $\succeq$ $\mbox{OMC}_{sl}$ $\succ$ ColorCrafter
\end{itemize}

In general, our study results confirm findings from Golebiowska and Coltekin~\cite{Golebiowska.2020} that multi-colored scales like OMC and Rainbow are more suitable for reading tasks while sequential scales like Viridis work better for comparing tasks.

Our design was developed on the application example of multidimensional meteorological cloud data. 
Our evaluation was data agnostic making our results extendable to other datasets.
The choice of color scheme however was informed by our collaboration with domain expert therefore, in the future, we would like to test the performance of our approach on other data types from various application areas, potentially exploring new color schemes. Furthermore, we have to investigate color blindness for our new design.

\acknowledgments{
The authors would like to thank all study participants and the reviewers, whose suggestions helped improve this paper.}

\bibliographystyle{abbrv-doi}

\bibliography{Literatur}
\end{document}